\documentclass[aps,prd,preprint,floatfix,nofootinbib,superscriptaddress,showpacs,amssymb]{revtex4}
 \usepackage{graphicx}
 \usepackage{type1cm}
 \def\deg{^\circ}
\def\deg{^\circ}

\def\Z0{${\em Z^0\/}$}

\def\r#1 {$^{#1}$}

\newcommand{\gevc} { {\rm GeV/}c}
\newcommand{\gevcc}{ {\rm GeV/}c^2}

\def\gepsfcentered#1{
  \def\testit{#1}
  \def\lbracket{[}
  \ifx\testit\lbracket
    \let\dofilecmd=\gepsfwithopt
  \else
    \let\dofilecmd=\gepsfnoopt
  \fi
  \dofilecmd}

\def\gepsfnoopt#1{
  \begin{center}
  \leavevmode
  \epsffile{#1}
  \end{center}}

\def\gepsfwithopt#1 #2 #3 #4]#5{
  \begin{center}
  \leavevmode
  \gepsfmaxx=0.94\textwidth
  \epsffile[#1 #2 #3 #4]{#5}
  \end{center}}

\newdimen\gepsfmaxx
\def\epsfsize#1#2{
  \ifnum \epsfxsize=0
    \ifnum \epsfysize=0
      \ifnum #1 > \gepsfmaxx
        \gepsfmaxx
      \else
        #1
      \fi
    \else
      \epsfxsize
    \fi
  \else
    \epsfxsize
  \fi
}

\begin{document}
 \title {An additional study of multi-muon events produced in
         {\boldmath $p\bar{p}$} collisions at {\boldmath $\sqrt{s}=1.96$} TeV 
}
 \affiliation{Institute of Physics, Academia Sinica, Taipei, Taiwan 11529, Republic of China} 
\affiliation{Argonne National Laboratory, Argonne, Illinois 60439, USA} 
\affiliation{University of Athens, 157 71 Athens, Greece} 
\affiliation{Baylor University, Waco, Texas 76798, USA} 
\affiliation{Istituto Nazionale di Fisica Nucleare Bologna, $^{aa}$University of Bologna, I-40127 Bologna, Italy} 
\affiliation{University of California, Davis, Davis, California 95616, USA} 
\affiliation{University of California, Los Angeles, Los Angeles, California 90024, USA} 
\affiliation{Instituto de Fisica de Cantabria, CSIC-University of Cantabria, 39005 Santander, Spain} 
\affiliation{Carnegie Mellon University, Pittsburgh, Pennsylvania 15213, USA} 
\affiliation{Enrico Fermi Institute, University of Chicago, Chicago, Illinois 60637, USA}
\affiliation{Comenius University, 842 48 Bratislava, Slovakia; Institute of Experimental Physics, 040 01 Kosice, Slovakia} 
\affiliation{Joint Institute for Nuclear Research, RU-141980 Dubna, Russia} 
\affiliation{Duke University, Durham, North Carolina 27708, USA} 
\affiliation{Fermi National Accelerator Laboratory, Batavia, Illinois 60510, USA} 
\affiliation{University of Florida, Gainesville, Florida 32611, USA} 
\affiliation{Laboratori Nazionali di Frascati, Istituto Nazionale di Fisica Nucleare, I-00044 Frascati, Italy} 
\affiliation{University of Geneva, CH-1211 Geneva 4, Switzerland} 
\affiliation{Glasgow University, Glasgow G12 8QQ, United Kingdom} 
\affiliation{Harvard University, Cambridge, Massachusetts 02138, USA} 
\affiliation{Division of High Energy Physics, Department of Physics, University of Helsinki and Helsinki Institute of Physics, FIN-00014, Helsinki, Finland} 
\affiliation{University of Illinois, Urbana, Illinois 61801, USA} 
\affiliation{The Johns Hopkins University, Baltimore, Maryland 21218, USA} 
\affiliation{Center for High Energy Physics: Kyungpook National University, Daegu 702-701, Korea; Seoul National University, Seoul 151-742, Korea; Sungkyunkwan University, Suwon 440-746, Korea; Korea Institute of Science and Technology Information, Daejeon 305-806, Korea; Chonnam National University, Gwangju 500-757, Korea; Chonbuk National University, Jeonju 561-756, Korea} 
\affiliation{Ernest Orlando Lawrence Berkeley National Laboratory, Berkeley, California 94720, USA} 
\affiliation{University of Liverpool, Liverpool L69 7ZE, United Kingdom} 
\affiliation{University College London, London WC1E 6BT, United Kingdom} 
\affiliation{Centro de Investigaciones Energeticas Medioambientales y Tecnologicas, E-28040 Madrid, Spain} 
\affiliation{Massachusetts Institute of Technology, Cambridge, Massachusetts 02139, USA} 
\affiliation{University of Michigan, Ann Arbor, Michigan 48109, USA} 
\affiliation{Michigan State University, East Lansing, Michigan 48824, USA}
\affiliation{Institution for Theoretical and Experimental Physics, ITEP, Moscow 117259, Russia}
\affiliation{University of New Mexico, Albuquerque, New Mexico 87131, USA} 
\affiliation{Northwestern University, Evanston, Illinois 60208, USA} 
\affiliation{The Ohio State University, Columbus, Ohio 43210, USA} 
\affiliation{Okayama University, Okayama 700-8530, Japan} 
\affiliation{Osaka City University, Osaka 588, Japan} 
\affiliation{University of Oxford, Oxford OX1 3RH, United Kingdom} 
\affiliation{Istituto Nazionale di Fisica Nucleare, Sezione di Padova-Trento, $^{bb}$University of Padova, I-35131 Padova, Italy} 
\affiliation{LPNHE, Universite Pierre et Marie Curie/IN2P3-CNRS, UMR7585, Paris, F-75252 France} 
\affiliation{Istituto Nazionale di Fisica Nucleare Pisa, $^{cc}$University of Pisa, $^{dd}$University of Siena and $^{ee}$Scuola Normale Superiore, I-56127 Pisa, Italy} 
\affiliation{University of Pittsburgh, Pittsburgh, Pennsylvania 15260, USA} 
\affiliation{Purdue University, West Lafayette, Indiana 47907, USA} 
\affiliation{University of Rochester, Rochester, New York 14627, USA} 
\affiliation{The Rockefeller University, New York, New York 10065, USA} 
\affiliation{Istituto Nazionale di Fisica Nucleare, Sezione di Roma 1, $^{ff}$Sapienza Universit\`{a} di Roma, I-00185 Roma, Italy} 
\affiliation{Rutgers University, Piscataway, New Jersey 08855, USA} 
\affiliation{Istituto Nazionale di Fisica Nucleare Trieste/Udine, I-34100 Trieste, $^{gg}$University of Udine, I-33100 Udine, Italy} 
\affiliation{University of Tsukuba, Tsukuba, Ibaraki 305, Japan} 
\affiliation{Tufts University, Medford, Massachusetts 02155, USA} 
\affiliation{University of Virginia, Charlottesville, Virginia 22906, USA}
\affiliation{Waseda University, Tokyo 169, Japan} 
\affiliation{Wayne State University, Detroit, Michigan 48201, USA} 
\affiliation{University of Wisconsin, Madison, Wisconsin 53706, USA} 
\affiliation{Yale University, New Haven, Connecticut 06520, USA} 
\author{T.~Aaltonen}
\affiliation{Division of High Energy Physics, Department of Physics, University of Helsinki and Helsinki Institute of Physics, FIN-00014, Helsinki, Finland}
\author{B.~\'{A}lvarez~Gonz\'{a}lez$^w$}
\affiliation{Instituto de Fisica de Cantabria, CSIC-University of Cantabria, 39005 Santander, Spain}
\author{S.~Amerio}
\affiliation{Istituto Nazionale di Fisica Nucleare, Sezione di Padova-Trento, $^{bb}$University of Padova, I-35131 Padova, Italy} 
\author{D.~Amidei}
\affiliation{University of Michigan, Ann Arbor, Michigan 48109, USA}
\author{A.~Anastassov}
\affiliation{Northwestern University, Evanston, Illinois 60208, USA}
\author{A.~Annovi}
\affiliation{Laboratori Nazionali di Frascati, Istituto Nazionale di Fisica Nucleare, I-00044 Frascati, Italy}
\author{J.~Antos}
\affiliation{Comenius University, 842 48 Bratislava, Slovakia; Institute of Experimental Physics, 040 01 Kosice, Slovakia}
\author{G.~Apollinari}
\affiliation{Fermi National Accelerator Laboratory, Batavia, Illinois 60510, USA}
\author{A.~Apresyan}
\affiliation{Purdue University, West Lafayette, Indiana 47907, USA}
\author{T.~Arisawa}
\affiliation{Waseda University, Tokyo 169, Japan}
% ...
\collaboration{CDF Collaboration\footnote{With visitors from $^a$University of MA Amherst,
Amherst, MA 01003, USA,
$^b$Istituto Nazionale di Fisica Nucleare, Sezione di Cagliari, 09042 Monserrato (Cagliari), Italy,
$^c$University of CA Irvine, Irvine, CA  92697, USA,
$^d$University of CA Santa Barbara, Santa Barbara, CA 93106, USA,
$^e$University of CA Santa Cruz, Santa Cruz, CA  95064, USA,
$^f$CERN,CH-1211 Geneva, Switzerland,
$^g$Cornell University, Ithaca, NY  14853, USA, 
$^h$University of Cyprus, Nicosia CY-1678, Cyprus, 
$^i$University College Dublin, Dublin 4, Ireland,
$^j$University of Fukui, Fukui City, Fukui Prefecture, Japan 910-0017,
$^k$Universidad Iberoamericana, Mexico D.F., Mexico,
$^l$Iowa State University, Ames, IA  50011, USA,
$^m$University of Iowa, Iowa City, IA  52242, USA,
$^n$Kinki University, Higashi-Osaka City, Japan 577-8502,
$^o$Kansas State University, Manhattan, KS 66506, USA,
$^p$University of Manchester, Manchester M13 9PL, United Kingdom,
$^q$Queen Mary, University of London, London, E1 4NS, United Kingdom,
$^r$University of Melbourne, Victoria 3010, Australia,
$^s$Muons, Inc., Batavia, IL 60510, USA,
$^t$Nagasaki Institute of Applied Science, Nagasaki, Japan, 
$^u$National Research Nuclear University, Moscow, Russia,
$^v$University of Notre Dame, Notre Dame, IN 46556, USA,
$^w$Universidad de Oviedo, E-33007 Oviedo, Spain, 
$^x$Texas Tech University, Lubbock, TX  79609, USA,
$^y$Universidad Tecnica Federico Santa Maria, 110v Valparaiso, Chile,
$^z$Yarmouk University, Irbid 211-63, Jordan,
$^{hh}$On leave from J.~Stefan Institute, Ljubljana, Slovenia, 
}}
\noaffiliation

 \begin{abstract}
 We present one additional study of  multi-muon events produced at the
 Fermilab Tevatron collider and recorded by the CDF~II detector. 
 We use a data  set acquired with a dedicated dimuon trigger and
 corresponding to an integrated luminosity of 3.9 fb$^{-1}$.
 We investigate the distribution of the azimuthal angle between the two
 trigger muons in events containing at least four additional muon
 candidates to test the compatibility of these events with originating
 from known QCD processes. We find that this distribution is markedly
 different from what is expected from such QCD processes and this 
 observation strongly disfavours the possibility that multi-muon events
 result from an underestimate of the rate of misidentified muons in
 ordinary QCD events.
\end{abstract}
 \preprint{FERMILAB-PUB-11-241-E}
 \maketitle
 This Letter reports on one additional test on the possible origin of
 multi-muon events observed at the Tevatron.
 These events were identified in a previous study~\cite{a0disc} of a
 data set acquired with two central ($|\eta|<0.7$) primary (or trigger)
 muons, each with transverse momentum $p_T \geq 3 \; \gevc$, and with
 invariant mass larger than 5 $\gevcc$ and smaller than 80 $\gevcc$.
 That study shows that many long-standing inconsistencies between 
 measured and predicted properties of the correlated $b\bar{b}$
 production and semileptonic decay at hadron
 colliders~\cite{2mucdf,d0b2,bmix,dilb} could be explained by the presence
 of a relevant source of muons which appear to be mostly produced beyond 
 the beam pipe of radius 1.5 cm (this contribution is whimsically referred
 to as ghost events because they were unnoticed or ignored by previous
 measurements).
 Within the large uncertainty of the prediction, mostly based on simulations,
 the observed rate of ghost events is found to be consistent with being
 produced by muons arising from in-flight-decays of pions and kaons,
 or punchthrough of hadronic prongs from $K_S^0$ or hyperon decays.
 However, a search in ghost events for additional muons with
 $p_T \geq 2 \; \gevc$ and $|\eta| \leq 1.1$ and contained in a 
 $\cos \theta \geq 0.8$ cone around the direction of a primary muon 
 selects a small but significant fraction of events with a large content
 of muon candidates that appears difficult to account for in terms of
 known sources with the present understanding of the CDF~II detector,
 trigger, and event reconstruction. 

 A more recent study by the CDF collaboration~\cite{impr-mumu} has improved
 the estimate of the contribution of ordinary sources to ghost events.
 This study addresses in particular the contribution from pion and kaon
 in-flight-decays. In  1426 pb$^{-1}$ of data, there are $54437 \pm 14171$
 ghost events and $12169 \pm 1319$ ghost events with three or more muons
 which cannot yet be accounted for with ordinary sources.
  
 In this Letter, we investigate the distribution of the azimuthal angle
 ($\delta \phi$) between the two primary muons in events in which both
 primary muons are accompanied by at least one (or two) additional muon
 candidates in a $\cos \theta \geq 0.8$ cone around their direction,
 and compare it to those for all QCD sources known to produce dimuon
 events: $b\bar{b}$, $c\bar{c}$, and $\Upsilon$ production or  events
 in which one trigger muon is due to hadrons misidentified as muons
 (cosmic rays are removed from the data sample and the contribution
 of secondary interactions in the detector volume is negligible~\cite{a0disc}).
 As discussed in Ref.~\cite{a0disc}, known QCD sources produce a handful
 of events with four and none with six muon candidates.
 However, if the unaccounted multi-muon events were generated by a gross
 underestimate of the number of additional muons mimicked by hadrons in
 ordinary QCD events, the  $\delta \phi$ distribution of primary muons
 in multi-muon events would be similar to that of ordinary QCD events
 in which the large contribution of next-to-leading order (NLO) terms
 due to initial and final state radiation results in a broader $\delta \phi$
 distribution than that predicted by the Born (LO) approximation.
 In fact, the $\delta \phi$ distribution of pairs of $b$ hadrons or
 jets is traditionally used to determine the relative contribution of
 NLO to LO terms~\cite{kevin}. This type of comparison was also suggested
 by Ref.~\cite{barb}, in which the excess of multi-muon events is modeled
 with the decay of two colorless particles produced through the exchange
 of a heavy object. In such a hypothetical case, their deviation from
 the back-to-back configuration in the azimuthal angle ($\delta \phi = \pi$)
 is only caused by initial state radiation of the incoming quarks and
 is expected to be small.

 The study presented here  uses a dimuon data set corresponding to an
 integrated luminosity of 3.9 fb$^{-1}$ and selected with the same
 requirements used in Ref.~\cite{a0disc}. High precision charged particle
 tracking is  provided by a large central drift chamber surrounding a trio
 of silicon tracking devices composed of eight layers of silicon microstrip
 detectors ranging in radius from $1.5$ to $28$~cm in the pseudorapidity
 region $|\eta|<1$~\cite{cdfdet}. The tracking detectors are inside a 
 1.4 T solenoid which in turn is surrounded by electromagnetic and hadronic
 calorimeters. Outside the calorimeters, drift chambers in the region
 $|\eta| \leq 1.1$ provide muon identification. We search events for
 additional muons using tracks with $p_T \geq 2 \; \gevc$ and $|\eta| \leq 1.1$.
 The rate of additional muons mimicked by hadronic punchthrough is 
 estimated with a probability per track derived by using kaons and pions
 from $D^{*\pm} \rightarrow \pi^\pm  D^0$ with $D^0 \rightarrow K^+ \pi^-$
 decays~\cite{a0disc,impr-mumu,bbxs}. The difference between observed
 additional muons and predicted misidentifications is referred to as real
 muons.

\begin{figure}[htb]
 \begin{center}
 \leavevmode
 \includegraphics*[width=0.5\textwidth]{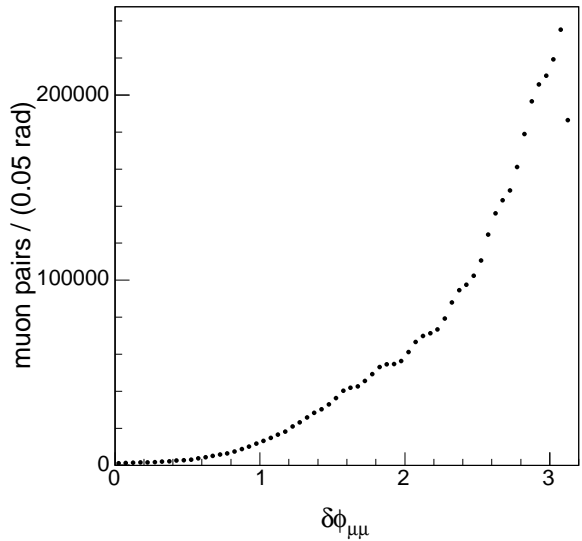}
 \caption[]{Distribution of the azimuthal angle $\delta \phi$ between the
            two trigger muons for all events.}
 \label{fig:fig_1}
 \end{center}
 \end{figure}

\begin{figure*}[htb]
 \begin{center}
 \leavevmode
 \includegraphics*[width=0.5\textwidth]{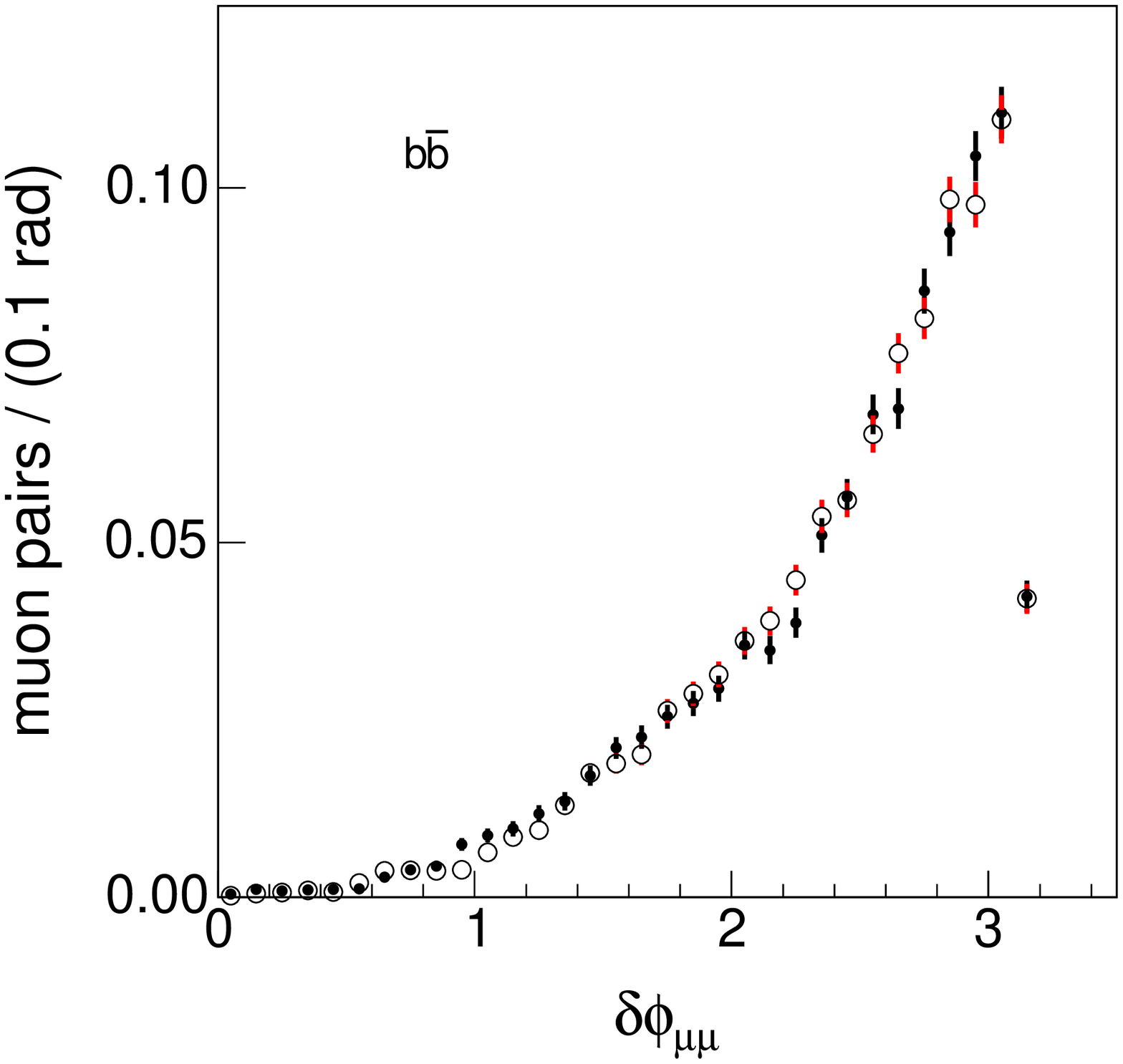}\includegraphics*[width=0.5\textwidth]{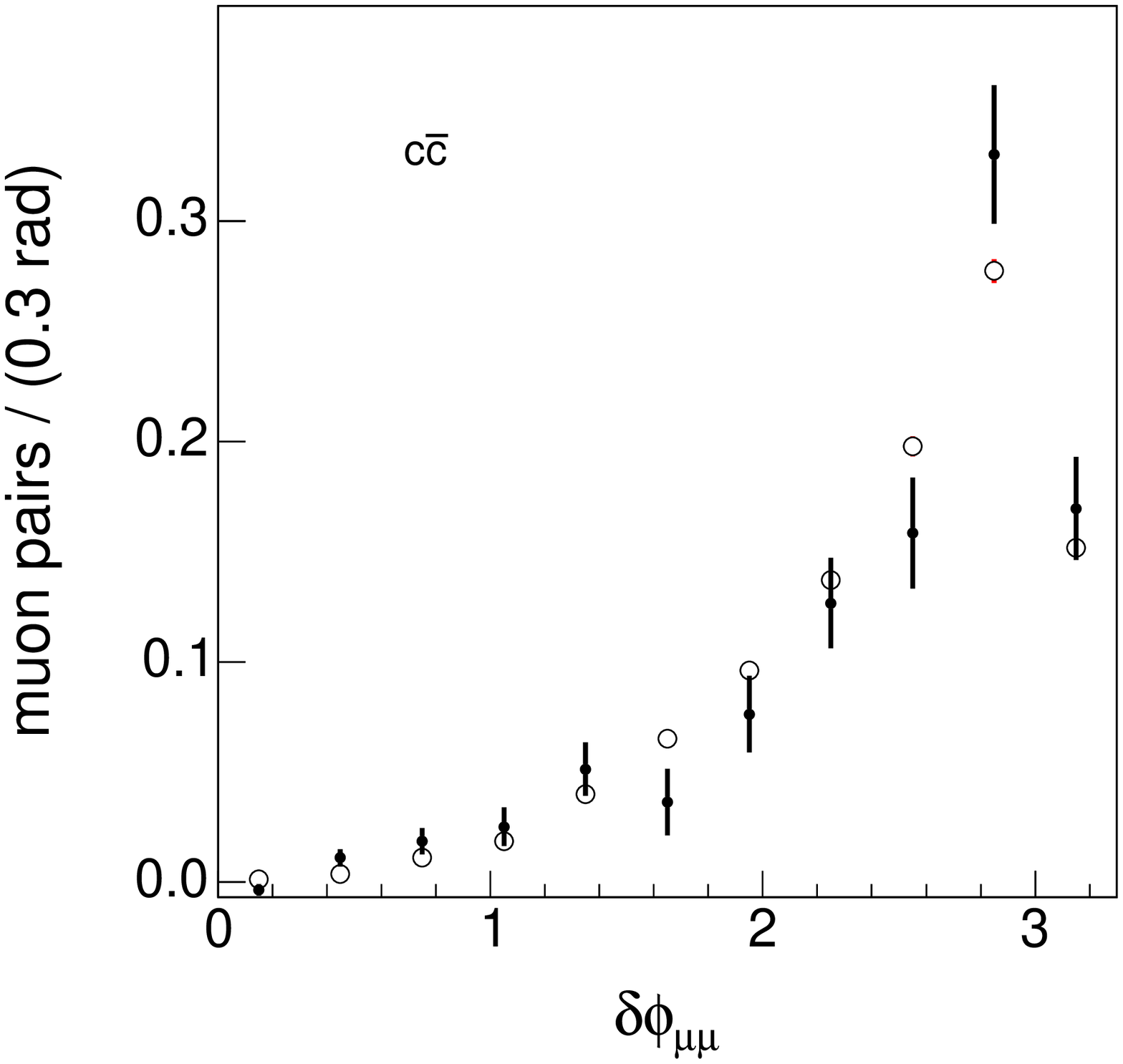}
 \caption[]{The distributions ($\bullet$) of the azimuthal angle $\delta \phi$
            between  trigger muons due to  (left) $b\bar{b}$ and (right)
            $c\bar{c}$ production are compared to the corresponding heavy
            flavor simulations ($\circ$). Distributions are normalized to
            unit area.}
 \label{fig:fig_2}
 \end{center}
 \end{figure*}

 The $\delta \phi$ distribution for all 3.9 M events is shown in
 Fig.~\ref{fig:fig_1}.
 Figure~\ref{fig:fig_2} compares to the corresponding heavy flavor 
 simulations the $\delta \phi$ distribution of trigger muons due to
 $b\bar{b}$ and $c\bar{c}$ production. This figure is reproduced from
 Ref.~\cite{bbxs} that has measured $\sigma_{b\rightarrow\mu,\bar{b}\rightarrow \mu}$
 and $\sigma_{c\rightarrow\mu,\bar{c}\rightarrow \mu}$ in a dimuon data set
 corresponding to a luminosity of 742 pb$^{-1}$.
 In the $b\bar{b}$ case, the distribution has an average of 2.5 with
 a rms deviation of 0.8 rad. The long and important tail extending to
 $\delta \phi=0$ is due to NLO terms and the non-perturbative fragmentation
 function of  $b$ quarks. In $c\bar{c}$ events, because of the smaller
 quark mass, NLO terms are approximately a factor of three larger and 
 the fragmentation function is much softer. Accordingly, the  $\delta \phi$ 
 distribution has a smaller average (2.4 rad) and a larger rms deviation
 (0.9 rad).

\begin{figure*}[htb]
 \begin{center}
 \leavevmode
 \includegraphics*[width=0.5\textwidth]{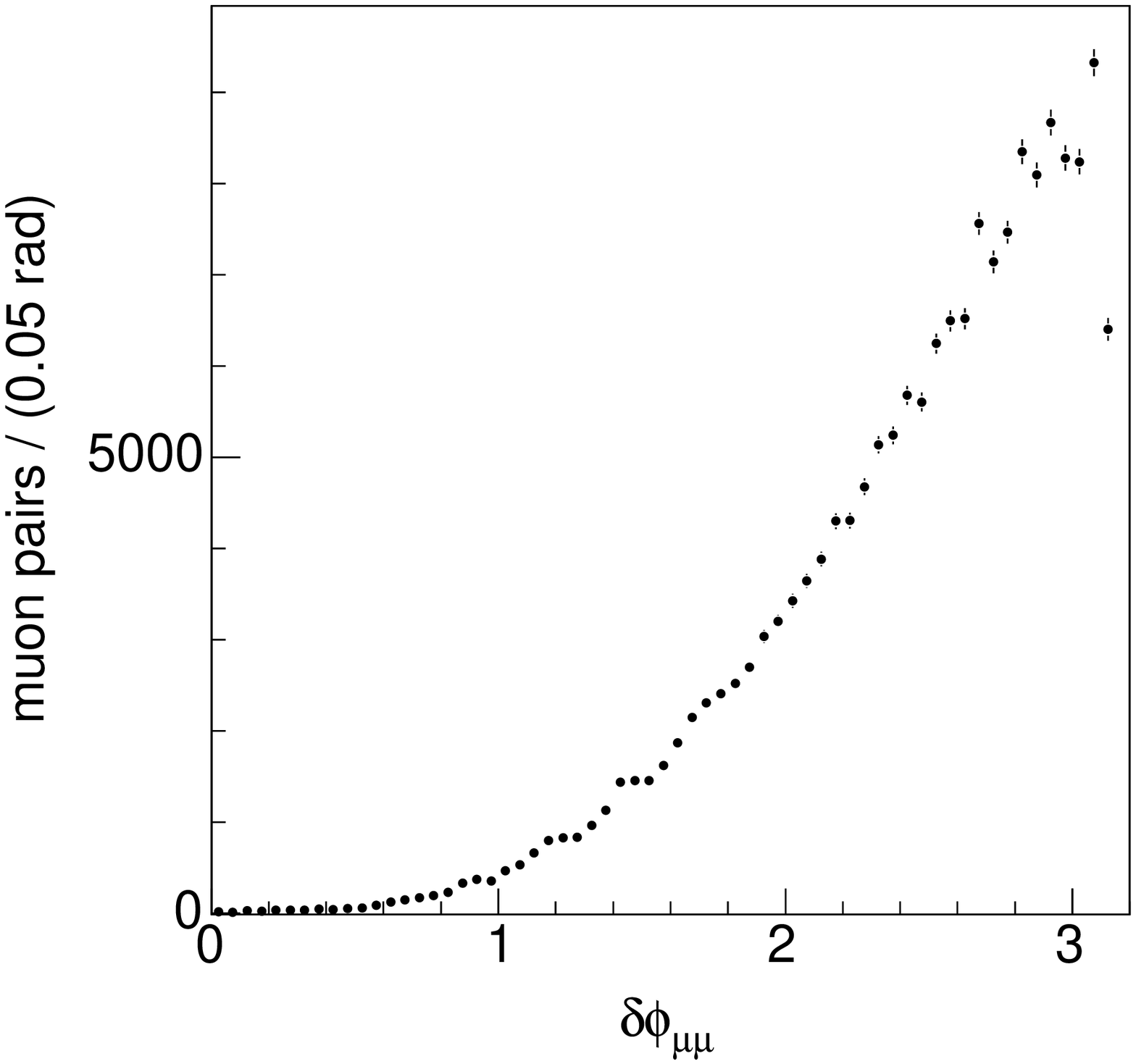}\includegraphics*[width=0.5\textwidth]{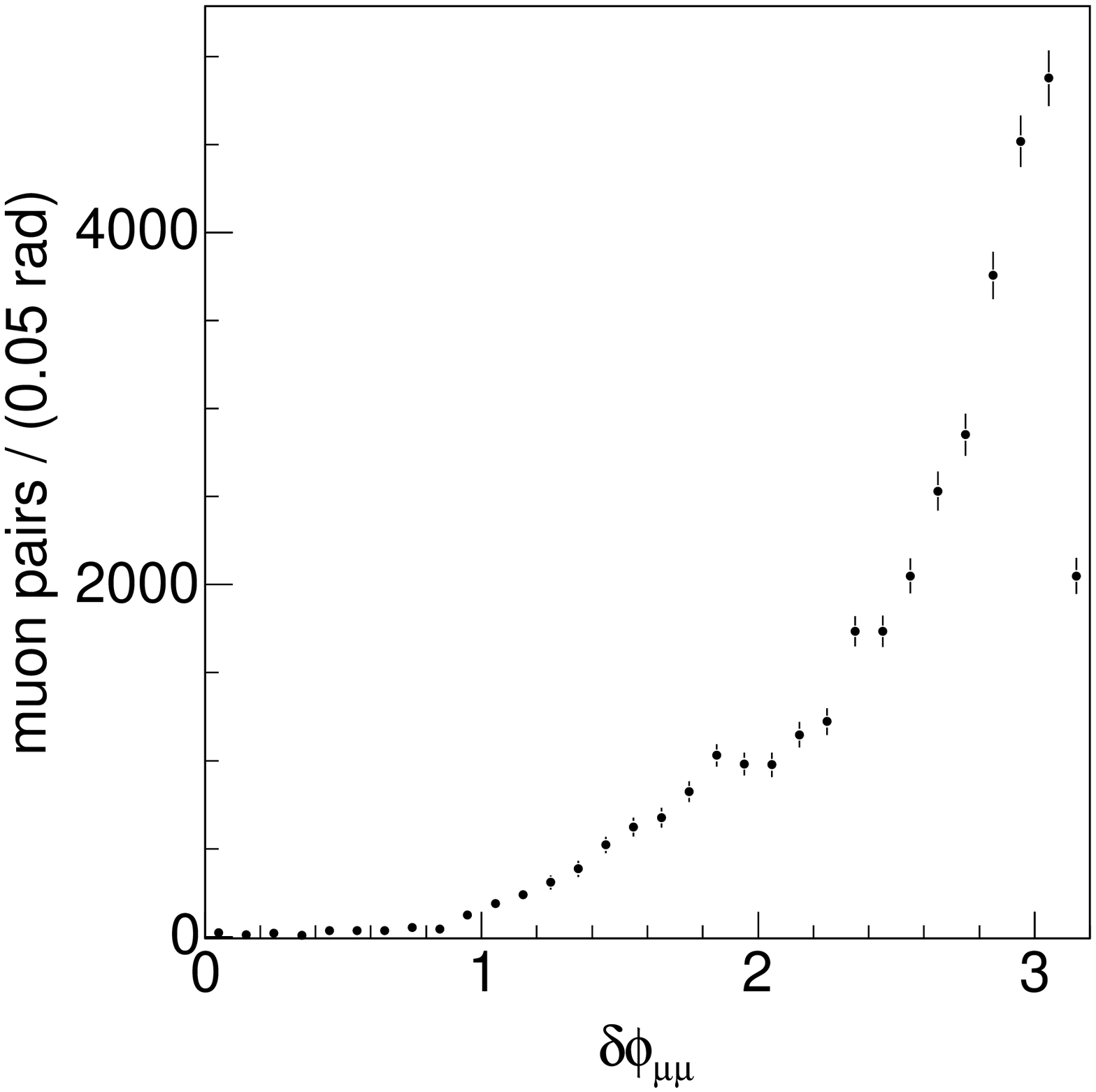}
 \caption[]{Distribution of the azimuthal angle $\delta \phi$ between the
            two trigger muons produced by $\Upsilon$ decays (left) and for
            events (right) in which one primary muon is mimicked by a pion
            produced by an identified  $K^0_S$ decay. The combinatorial
            background underneath the $\Upsilon$ and $K^0_S$ signals has been
            removed with a sideband subtraction method.}
 \label{fig:fig_3}
 \end{center}
 \end{figure*}

 The azimuthal-angle distribution for primary muons produced by $\Upsilon(1S)$
 decays is expected to be similar to those for heavy flavors because the
 final state contains  a bleaching gluon recoiling against the $\Upsilon$ meson.
 This distribution, shown in Fig.~\ref{fig:fig_3}, is constructed using
 muon pairs with invariant mass in the range $9.28-9.6 \; \gevcc$.
 As in Ref.~\cite{bbxs}, the combinatorial background under the $\Upsilon(1S)$
 signal is removed with a sideband subtraction technique.
 A similar $\delta \phi$ distribution is also expected for those cases in
 which one muon is mimicked by a track in the jet recoiling against a muon
 due to a heavy-quark semileptonic decay.
 Figure~\ref{fig:fig_3}  shows  the  $\delta \phi$ distribution of primary
 muons when one of them is mimicked by pions produced by $K^0_S$ decays.
 As in Ref.~\cite{impr-mumu}, we select $K^0_S \rightarrow \pi^+ \pi^-$
 with a $\pi \rightarrow \mu$ misidentification by combining primary muons
 with tracks of opposite charge and $p_T \geq 0.5\; \gevc$. We select pairs
 consistent to those arising from a common three-dimensional vertex.
 We also take advantage of the $K^0_S$ long lifetime to suppress the
 combinatorial background.  We further require that the distance between
 the $K^0_S$ vertex and the event primary vertex, corrected by the $K^0_S$
 Lorentz boost, corresponds to  $ct>0.1$ cm. We select  $K^0_S$ candidates
 with invariant mass in the range $0.47-0.52\; \gevcc$
 (see Fig.~3 of Ref.~\cite{impr-mumu}), and remove the combinatorial
 background with a sideband subtraction technique.  
 
 In summary, the $\delta \phi$ distributions of primary muons produced
 by known QCD processes peak at $\delta \phi \simeq \pi$, and exhibit
 a significant tail extending to  $\delta \phi=0$. Depending on the
 production mechanism, the mean and rms deviation of these distributions 
 are in the range of $2.4-2.5$ rad and  $0.7-0.9$ rad, respectively.

\begin{figure*}[htb]
 \begin{center}
 \leavevmode
 \includegraphics*[width=\textwidth]{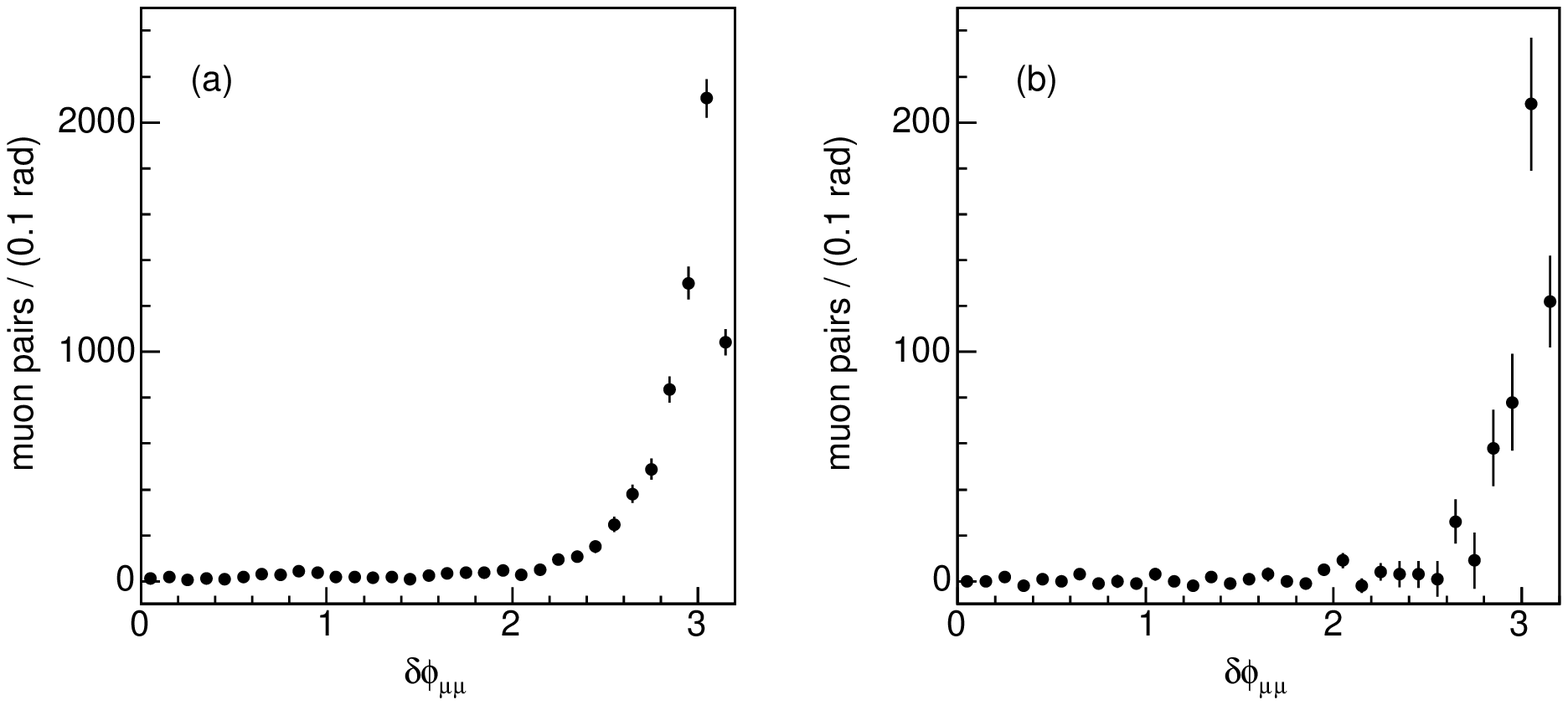}
 \caption[]{Distribution of the azimuthal angle $\delta \phi$ between the
            two trigger muons accompanied by at least (a) one or (b) two
            additional real muons in a 36.8$^{\deg}$ cone around their
            direction.}
 \label{fig:fig_4}
 \end{center}
 \end{figure*}

 The $\delta \phi$ distributions in the subset of events in which each 
 trigger muon is accompanied by at least one or at least two additional
 real muons are shown in Fig.~\ref{fig:fig_4}. These $\delta \phi$
 distributions, with mean of 2.9 rad and rms deviation of 0.2 rad and
 without any tail below  $\delta \phi=2.5$ rad, are different from those
 of primary muons due to all known QCD sources.

 In conclusion, as mentioned earlier, within our present understanding
 of the CDF-detector response no known sources produce events in which
 each $\cos \theta \geq 0.8$ angular cone around a primary muon contain at
 least two additional real muons. Had the additional muons been produced
 by a subtle failure of our method to evaluate the fake-muon contribution,
 the resulting $\delta \phi$ distribution of primary muons would have been
 found consistent with those typical of ordinary QCD processes.  

  We thank the Fermilab staff and the technical staffs of the participating
  institutions for their vital contributions. This work was supported by the
  U.S. Department of Energy and National Science Foundation; 
  the Italian Istituto Nazionale di Fisica Nucleare; the Ministry of Education,
  Culture, Sports, Science and Technology of Japan;  the National Science
  Council of the Republic of China; the Swiss National Science Foundation;
  the A.P. Sloan  Foundation; 
  the Korean Science and Engineering Foundation and the Korean Research 
  Foundation; the Science and Technology Facilities Council and the Royal
  Society, UK;  the Institut National de Physique Nucleaire et Physique
  des Particules/CNRS; the Russian Foundation for Basic Research; 
  the Ministerio de Ciencia e Innovaci\'{o}n, Spain; the European Community's
  Human Potential Programme; the Slovak R\&D Agency;
  and the Academy of Finland.

 \end{document}